\documentclass[12pt]{article}

\newenvironment{sciabstract}{%
\begin{quote} \bf}
{\end{quote}}

\title{Large anomalous Hall effect in a silicon-based magnetic semiconductor}

\author{Ncholu Manyala$^{1,4}$, Yvan Sidis$^{1\dag}$, John F. DiTusa$^{1\ast}$, Gabriel Aeppli$^{2}$,\\
David P. Young$^{1}$, and Zachary Fisk$^{3}$\\
\\
\normalsize{$^{1}$Department of Physics and Astronomy,}\\
\normalsize{Louisiana State University, Baton Rouge Louisiana 70803, USA}\\
\normalsize{$^{2}$London Centre for Nanotechnology and Department of Physics and Astronomy,}\\
\normalsize{UCL, London WC1E 6BT, UK}\\
\normalsize{$^{3}$National High Magnetic Field Facility,}\\
\normalsize{Florida State University, Tallahassee, Florida 32306, USA}\\
\normalsize{$^{4}$Department of Physics and Electronics,}\\
\normalsize{National University of Lesotho, P.O.\ Roma 180, Maseru 100, Lesotho}\\
\\
\normalsize{$^{\dag}$Present address Laboratoire L\'{e}on Brillouin,}\\
\normalsize{CEA-CNRS, CE-Saclay, 91911 Gif-Sur-Yvette, France}\\
}

\date{}

\begin{document} 


\baselineskip24pt


\maketitle


\begin{sciabstract}
Magnetic semiconductors are attracting high interest because of their potential use for 
spintronics,
a new technology which merges electronics and manipulation of conduction electron spins. 
(GaMn)As and (GaMn)N have recently emerged as the most popular materials for this new 
technology. While Curie temperatures are rising towards room temperature, these 
materials can only be fabricated in thin film form, are heavily defective, and are not 
obviously compatible with Si. We show here that it is productive to consider transition 
metal monosilicides as potential alternatives. In particular, we report the discovery 
that the bulk metallic magnets derived from doping the narrow gap insulator FeSi with 
Co share the very high anomalous Hall conductance of (GaMn)As, while displaying Curie temperatures 
as high as 53 K. Our work opens 
up a new arena for spintronics, involving a bulk material based only on transition metals 
and Si, and which we have proven to display a variety of large magnetic field effects on 
easily measured electrical properties. 
%
%
\end{sciabstract}


\section*{Introduction}
Magnetic semiconductors are attracting interest because they are more likely than ordinary 
metals to serve as injectors for spintronics applications.
 Recent successes include the 
discoveries that GaAs and GaN can display ferromagnetism on substitution of Ga by 
Mn$^{1-4}$. 
GaAs and GaN are semiconductors with important uses in electro-optics, 
and the magnetism is clearly derived from the Mn 
ions while the carriers are derived from other dopants.
Recent work has shown that the 
narrow gap$^{5,6}$ semiconductor FeSi can be doped via substitution of a single species, Co, 
for Fe to produce a low carrier density ($n$) magnetic metal with exceptional
magnetoconductance (MC)$^7$. That this is a truly itinerant low $n$
system, in contrast to various semiconductors with Mn substitution, is apparent both 
from the absence of magnetism in the isostructural end members FeSi and CoSi of the 
Fe$_{1-y}$Co$_y$Si dilution series, the fact that each Co atom adds one Bohr magneton to the magnetic 
polarization for low doping, and exceptional MC entirely understandable in 
terms of strong exchange enhancement of a disordered metal with Coulomb interactions$^{7,8}$. 
In 
the present paper, we report the discovery of a large anomalous Hall effect (AHE) in addition to 
the other unusual properties already found for Fe$_{1-y}$Co$_y$Si. This is an important 
result because it demonstrates that large AHE may be a general feature of magnetic semiconductors, and
in particular can be found in 
low $n$, half-metallic materials with significant spin-orbit (SO) coupling evident in their band 
structures. Our measurements show that these diverse requirements can be achieved 
by simple chemical substitutions into a compound of two very common elements, iron and silicon.

\section{Monosilicides}
      The monosilicides, FeSi, CoSi, and MnSi all have the same cubic B-20
crystal structure allowing the exploration of Fe$_{1-x}$Mn$_x$Si
and Fe$_{1-y}$Co$_y$Si for all $x$ and $y$ between 0 and 1 (see 
Fig.\ 1)$^9$. FeSi is fascinating in itself as a 'Kondo insulator', 
the designation for the insulating parents of the heavy fermion (HF) compounds.
It transmits light in the far infra-red, with an optical gap of 
60 meV, and originally attracted attention because, surprisingly, it has a response 
to external magnetic field which is large at room temperature but vanishes as the temperature
approaches zero$^{5,6,10,11}$. 
When doped with holes (Mn or Al) or electrons (Co) an 
insulator to metal (MI) transition occurs at a doping level of $\sim 
0.01$ (see Fig.\ 1)$^{7,12,13}$. 
While electron doping beyond the MI transition almost immediately produces 
a helimagnetic (HM) ground state, hole doping produces a simple 
paramagnetic (PM) metal$^{7,8,12}$. 

     MnSi has long been known$^{14}$ as a classic weak itinerant ferromagnet (FM), 
which continues to provide surprises in the area of metal physics$^{15,16}$. 
Our Fe$_{1-x}$Mn$_{x}$Si samples with $x\le 0.8$ remain 
PM down to the lowest temperature ($T$) measured (1.7 K).
In this article we focus on the HM phases and in particular on Fe$_{1-y}$Co$_y$Si samples 
which are close in composition to the anomalous insulator FeSi. In this range of Co concentration,
$y \le 0.3$, magnetization measurements above 2 kG along with high field Hall effect measurements
reveal that each Co dopant adds one Bohr magneton to the magnetic polarization$^7$ and one electron 
carrier to FeSi (Fig.\ 1d). We have concluded from these data that the electron gas in these 
highly itinerant magnets is fully polarized at low temperature$^7$.
This means that by examining the transport and magnetic 
behavior across 
the Fe$_{1-x,y}$Mn$_x$Co$_y$Si series, we can study the continuous evolution
from a classic weak itinerant magnet, to a metallic paramagnet,  to a Kondo (or strongly 
correlated) insulator, and
finally a fully polarized itinerant magnetic metal$^{7,9,12}$ 
all without a change in crystal structure (see Fig.\ 1).

\section{Experimental Results}
	Our main result of a large Hall resistivity ($\rho_{xy}$) in Co doped FeSi can be seen 
in Figs.\ 2 and 3. In Fig.\ 2 the magnetic field ($H$) dependence of $\rho_{xy}$ of
a few of our HM samples is displayed. As is common for ferromagnets and strong paramagnets,
the Hall effect
has two contributions, one proportional to $H$, which is 
the ordinary term from which we extract the carrier densities, and 
the second determined by $M(H)$ $^{17,18}$.  
To highlight this second contribution it is
customary to write $\rho_{xy}$ as $\rho_{xy} = R_0 H + 4 \pi M R_S$.
Here $R_0$ is the Hall effect resulting from the Lorentz force on the carriers in 
the same manner as in PM materials and $R_S$ is referred to as the
anomalous Hall constant. Thus in HMs and FMs at 
$T$ below the Curie temperature ($T_c$), $\rho_{xy}$ has roughly the same $H$
dependence as the magnetization ($M$)$^{17,18}$.
This feature is demonstrated for Fe$_{1-y}$Co$_y$Si in Figs.\ 2a and 3a where
$\rho_{xy}$ has a large linear field dependence below 2 kG, the field where M(H) saturates  
in Fig.\ 2c. Beyond 2 kG
$\rho_{xy}$ becomes much less field--dependent (Fig.\ 3a). At these high fields 
$\rho_{xy}$ takes on the usual dependence on $n$
and $H$, $d\rho_{xy} / dH = R_0 = 1 / nec $ in its simplest form. It is for magnetic fields
less than 2 kG that $\rho_{xy}$ is proportional
to $M$ and the anomalous contribution dominates$^{17,18}$.

	For comparison we have also plotted $\rho_{xy}$ and $M$ for our MnSi and 
Fe$_{0.1}$Mn$_{0.9}$Si sample in Fig.\ 2. It is apparent from the figure that 
although $M$ is of the same order and has a comparable
$H$ dependence, the Hall effect is vastly different for 
the Mn-rich and Co-containing compounds. 
In fact there is a difference of a factor of 150 between the low $H$
$\rho_{xy}$ of the Fe$_{0.9}$Co$_{0.1}$Si and Fe$_{0.1}$Mn$_{0.9}$Si samples. 
We have chosen to compare these two samples in detail since they have the same 
crystal structure, the same level of chemical substitution, and
HM ground states with $T_c = 10$ K. 

	A further comparison is shown in Fig.\ 2 b and d
which presents the temperature dependence of the zero $H$ resistivity ($\rho_{xx}$) and magnetic 
susceptibility ($\chi$) at 50 G. Again, after the differences in Curie points have been taken 
into account, the magnetic properties of 
these samples 
appear to be similar. At the same time the temperature 
dependence of the resistivity
is very different, $\rho_{xx}$ rises above its trend line on cooling
below $T_c$ for Co-doped FeSi while it drops below the trend line as spin disorder
scattering disappears for MnSi and Fe-doped MnSi. This effect has been discussed 
previously$^7$ in the context of greatly enhanced quantum interference 
effects in Fe$_{1-y}$Co$_y$Si. What interests us here is that our Fe$_{0.9}$Co$_{0.1}$Si
sample is 9 times more resistive
than the  Fe$_{0.1}$Mn$_{0.9}$Si sample and nearly 20 times more resistive
than MnSi. The differences in the two disordered alloys can be entirely accounted for
in terms of the Drude model where the low $T$ Hall mobility  ($\mu_H = R_0 / \rho_{xx}$)
is essentially unchanged as a function of $x$ and $y$ (see Fig.\ 1e), and
the carrier density is simply obtained from counting the surplus or deficit of
valence electrons introduced by chemical substitution into the insulating FeSi parent
compound. Perhaps this paradigm breaks down in the limit of pure MnSi, for which the
mobility is larger by a factor of two, but, given the
complexity of the band structure of the transition metal 
silicides$^{19}$, applies over a remarkably wide range of $x$ and $y$, as does the 
tendency of $R_0$ to reflect the simple electron/hole counts associated with the 
chemical substitution (Fig.\ 1d). Given that the doped Kondo insulator Fe$_{1-y}$Co$_y$Si
behaves differently from the classic, high carrier density, itinerant manganese-rich 
compounds, we will compare our data to those for the classic ferromagnetic 
semiconductor, (GaMn)As, below.

	To understand the origin of our results for $\rho_{xy}$, 
we now focus on the anomalous contribution to the Hall effect ($4\pi MR_S$),
especially in the ordered state. The main part of Fig.\ 3a shows $\rho_{xy}$ versus
applied field on a scale strongly expanded relative to that of Fig.\ 2a. 
Since the saturation magnetization ($M_S$) of 
the Fe$_{0.1}$Mn$_{0.9}$Si sample is $\sim 3.5$ times larger than $M_S$ of our
Fe$_{0.9}$Co$_{0.1}$Si sample, $R_S$ of these two compounds differ by a factor of 
$\sim 500$. The large Hall resistivity in Fe$_{1-y}$Co$_y$Si is therefore predominantly 
controlled by the anomalous Hall coefficient. Fig.\ 3b shows how the anomalous term depends 
on temperature for some of
our samples as well as (GaMn)As$^{2,3}$ with a $T_c$ of 110 K. Although there is a 
difference of a factor of $\sim 25$ between $R_S$ of Fe$_{1-y}$Co$_y$Si and (GaMn)As, the
temperature dependence of $R_S$ in Fe$_{1-y}$Co$_y$Si
resembles that of (GaMn)As much more closely than those of the Mn-rich
silicides. In particular, the data for Fe$_{1-y}$Co$_y$Si and (GaMn)As have only slight 
temperature dependencies and converge
on finite values as $T$ goes to zero, while for the Mn-rich samples, $R_S$ has a dramatic
temperature dependence falling quickly toward 
zero as $T$ goes to zero. We are thus left with a full Hall effect whose magnitude and temperature
dependence is a much more 
sensitive probe of whether we are dealing with a magnetic semiconductor such
as Fe$_{1-y}$Co$_y$Si rather than an itinerant magnet like MnSi.

\section{Discussion and Analysis}
     The commonly accepted theory of the AHE
relies on SO coupling between the carrier and the lattice
which produces a left-right asymmetry in the scattering$^{17,18}$. Above
$T_c$ the randomization of the spins leads to an insignificant
transverse electric field ($E_{y}$). However, a large $E_{y}$
results when the material has a non-zero $M$ due
to the alignment of the carrier spins. The alignment 
creates an abundance of scattering in one particular direction, and a net current
perpendicular to the longitudinal electric field. Thus an $E_y$ many times larger 
than that due to the 
Lorentz force is necessary to cancel this anomalous current.
The usual description of $R_S$ sums two contributions proportional to 
$\rho_{xx}$ and $\rho_{xx}^2$ known as 
the ``skew scattering'' and ``side-jump'' terms respectively$^{17,18}$. 
Since $\rho_{xx}$ of our $x=0.9$ and $y=0.1$ samples differ by only a factor of 
$\sim 9$, this theory cannot account for the difference in $R_S$ that we measure
unless we posit that even though the scattering rates in these two materials are similar, the
scattering in Fe$_{0.9}$Co$_{0.1}$Si is much more effective in producing
a perpendicular current.  


      What might account for the unusually large anomalous Hall effects for Fe$_{1-y}$Co$_y$Si, 
while leaving small values for isostructural MnSi? To explore this issue, it turns out to be
useful to place our discoveries in a broader context. Fig.\ 4 compares 
the Hall effects for $H=0.1$ T for
Fe$_{1-y}$Co$_y$Si and Fe$_{1-x}$Mn$_x$Si to those for a wide range of other materials.
As is well known and apparent in the figure, very large Hall
effects result from making semiconductors intrinsic and thus reducing $n$. 
However, unlike the semiconductors, magnetic materials
have large $\rho_{xy}$ (as much as a few $\mu \Omega$ cm) while
retaining both metallic $n$, and $\rho_{xx}$. In fact our 
Fe$_{1-y}$Co$_y$Si samples have $\rho_{xy}$ similar to nonmagnetic 
semiconductors with a factor of 250 times smaller $n$, while maintaining 
$\rho_{xx}$ at a level 5 to 20 times smaller than these very clean crystalline
semiconductors. (GaMn)As films in particular stand out as having very large Hall 
resistivities due in part to their
small carrier concentrations ($n \sim 1.5\times10^{20}$ cm$^{-3}$)$^2$. 
What is also apparent is that
our Fe$_{1-y}$Co$_y$Si samples have among the largest $\rho_{xy}$ 
measured at 4K for metallic (poly)crystalline FMs with moderate $n$.

	Apart from going to low $n$, a second route to large Hall effects that enhances 
$R_S$ is 
typically achieved in HF systems$^{20,21}$.   
Fig.\ 4c, where we plot $R_S$ against $\rho_{xx}$, 
makes clear that Fe$_{1-y}$Co$_y$Si is comparable to other HF and mixed valent systems, 
with much higher $n$. However, what sets Fe$_{1-y}$Co$_{y}$Si apart is that because 
it is a long-period HM rather than a PM, the field -induced magnetizations are 
much higher than for the PM HF compounds. 

    What can be distilled from the first two frames of Fig.\ 4 is that Fe$_{1-y}$Co$_y$Si follows 
the standard trend-lines of decreasing $\rho_{xy}$ with $n$, but is shifted from 
the main line (of slope 1) describing ordinary semiconductors as well as (surprisingly) MnSi 
by a factor combining the high low field $M$ of other FMs and an $R_S$ as large as that of 
the HF systems.  

   Fig.\ 4c allows us to compare the anomalous Hall constant with other itinerant magnets as well.
The upper half of this figure includes 
many of the half metallic materials, those with spin polarized Fermi gases, which have 
received a great deal of recent attention. This category includes the colossal magnetoresistive 
manganites$^{22}$, the half-Heusler materials$^{23}$, 
Sr$_2$FeMoO$_6$ $^{24}$ and Fe$_{1-y}$Co$_y$Si$^7$. At the same time a large 
number of magnetic materials roughly follow the $R_S \propto \rho_{xx}^2$ law indicated by
the red line in the figure.

      For disordered materials, including the magnetic semiconductors of interest here, the second
order term, or side-jump scattering, a virtual transition resulting in a 
transverse offset of the scattered wave functions, should dominate the AHE. 
However, as Jungwirth, Niu, and MacDonald 
point out$^{25}$ the theory of the side-jump term in $R_S$, originally introduced 
by Luttinger$^{26}$, does not depend on scattering to produce an effect. Instead, 
the AHE results 
from the change in the wave packet group velocity that occurs when electric fields are 
applied to a FM. As such it relies on the SO coupling inherent to the Bloch 
wave functions instead of the SO coupling to impurities or defects. Thus, it 
is a ground state property of the system and may account for the survival of 
the AHE down to low temperatures in materials 
such as (GaMn)As and Fe$_{1-y}$Co$_y$Si (see Fig.\ 3)$^{2,3}$. 
The relevant intrinsic quantity is the off--diagonal conductivity ($\sigma_{xy}$)
and not the Hall resistance which also includes extrinsic scattering terms. 
We therefore plot (see Fig.\ 5) $\sigma_{xy} = \rho_{xy} / \rho_{xx}^2$
as a function of $M$ for the same materials as in Fig.\ 4. Plotting the data in this 
way rather than the standard methods of Fig.\ 4, which are appropriate for extrinsic scattering
dominated off--diagonal conductivities, gives a new perspective on the different 
classes of materials.
There are now three separate regions of behavior occupied by the HF
materials (upper left corner), the carrier hopping systems (lower right corner), and the 
itinerant magnets which includes the magnetic semiconductors. One valuable insight
which is gained immediately is that the colossal magnetoresistive manganites, in the 
carrier hopping region of the diagram, are truly distinct from the itinerant magnets.

     The clear separation and general trends apparent in Fig.\ 5 suggest that such a plot 
can be highly valuable in characterizing the Hall effect mechanism and the importance of
SO coupling to the carrier transport. Beyond occupying different regions of the diagram,
there are very different trends for $\sigma_{xy}$ versus $M$. In particular, the itinerant 
magnets show an obvious, monotonic increase in $\sigma_{xy}$ with $M$, while the HF metals
display a very dramatic rise in $\sigma_{xy}$ with decreasing $M$; the general trend for 
the hopping systems is not so obvious$^{22,25,28}$. 
For the HF metals, the behaviour is consistent with
the $1 /M^3$ law suggested, as spelled out in the figure caption, by a combination of 
two well known empirical facts about HFs, namely the Kadowaki-Woods relation and a 
constant Wilson ratio$^{27}$.

	The itinerant magnets, including the magnetic semiconductors 
display a rough $\sigma_{xy} \sim M$ 
dependence which is especially clear for Fe$_{1-y}$Co$_y$Si where all available data over a wide
range of $H$ and $T$ are shown. Variations from this simple dependence evident in the data 
can be interpreted as a measure of the strength of the SO coupling from band structure 
effects$^{25}$. Probably the most 
striking result is that the anomalous $\sigma_{xy}$ in the HM monosilicides and the
Mn doped III-V semiconductors shown in Fig.\ 5 are of comparable magnitude. Fig.\ 5
and Fig.\ 2c 
highlight the similarity of $M$ of (GaMn)As and Fe$_{1-y}$Co$_y$Si at low temperature
despite the differences in the mechanisms producing the ferromagnetism. There are 
also large differences in $n$, $1.5\times 10^{20}$ cm$^{-3}$ in (GaMn)As and 
$4.4\times 10^{21}$ to $1.3\times 10^{22}$ cm$^{-3}$ in Fe$_{1-y}$Co$_y$Si, and masses, 
$m^*=0.5 m_e$ and $0.08 m_e$ for the heavy and light holes in (GaMn)As and $m^* = 30 m_e$ in
Fe$_{1-y}$Co$_y$Si$^{12,13}$. However, the theory of Ref.\ ({\it 25\/}) 
predicts that 
$\sigma_{xy}\propto m^* / n^{1/3}$, suggesting a trade-off between $m^*$ and $n$ that
creates comparable $\sigma_{xy}$ in these two FM semiconductors. 

	Fig.\ 5 also emphasizes that the differences between Fe$_{1-y}$Co$_y$Si and MnSi 
are as profound as are the similarities of Fe$_{1-y}$Co$_y$Si to (GaMn)As. Not only 
is the 5 K value an order of magnitude lower for MnSi, 
but $\sigma_{xy}$ falls rather than rises as M approaches saturation on cooling.

\section{Conclusions}
	To close, our work has five significant aspects. First, we have discovered
a strongly correlated metal with a very large anomalous Hall effect. Second, we demonstrate
that the large effect is most likely intrinsic - derived from band structure effects 
rather than due to impurity scattering. Third, the effect is not found 
for the isostructural MnSi, thus 
adding another$^{7}$ sharp distinction between classic weak itinerant 
ferromagnetism and semiconducting 
ferromagnets with nearly the same ordered moment and Curie temperature. 
Fourth,  our survey of the Hall 
effect in a wide
variety of materials of high current interest reveals that it can be more productive to 
look at the Hall conductivity than at the Hall resistance.  Finally, our observation 
of the similarity of $\sigma_{xy}$ in Fe$_{1-y}$Co$_y$Si
and (GaMn)As is another indication that doped Kondo insulators might 
be useful for spintronics and provides a strong point of contact between two major areas, namely 
magnetic semiconductors and strongly interacting Fermions.
What makes FeSi especially attractive is that it is completely miscible with CoSi and MnSi 
and produces FMs with $T_c$'s 
as large as 60 K$^{8}$, comparable to all but the highest $T_c$s of the variety of Mn doped III-V 
and II-VI semiconductors. 
There is now strong 
incentive to discover a doped Kondo insulator exhibiting room temperature ferromagnetism. 

\section{Methods}
 Samples were either polycrystalline pellets or 
small single crystals grown from Sb and Sn fluxes. We produced the polycrystalline pellets 
by arc melting high-purity starting materials in an argon atmosphere. 
To improve the sample homogeneity the resulting Fe$_{1-y}$Co$_y$Si 
(Fe$_{1-x}$Mn$_x$Si) samples were annealed  for 24 hrs.\ at 
1200 $^o$C (four days at 1000 $^oC$) in evacuated quartz 
ampoules. Powder x-ray spectra showed all samples to be single phase with a 
lattice
constant linearly dependent on Co and Mn concentration (Fig.\ 1(c)).
The linearity demonstrates that Co or Mn successfully replaces Fe over the entire 
concentration range ($0\le x \le 1$, $0\le y\le 1$). Energy dispersive x-ray 
microanalysis yielded
results consistent with the nominal concentrations. The electrical conductivity
and Hall effect were measured on rectangular samples with thin Pt wires attached with 
silver paste at 19 Hz using lock-in techniques. Finally, the magnetization measurements were
made between 1.7 and 400 K in magnetic fields between -5.0 and 5.0 T in a SQUID magnetometer
and from 0 to 32 T in a vibrating reed magnetometer at the NHMFL.

We thank D.A.\ Browne and J.Y.\ Chan for 
discussions. JFD, ZF, and GA acknowledge the support of NSF under
contract No.s DMR 0103892, DMR 0203214, and a Wolfson-Royal Society Research Merit Award, respectively.

$^\ast$To whom correspondence should be addressed; E-mail:  ditusa@rouge.phys.lsu.edu.

\begin{quote}

\begin{enumerate}

\item Ohno, H., Munekata, H., Penney, T., von Molnar, S. \& Chang, L.L. Magnetotransport
properties of p-type (In,Mn)As diluted magnetic III-V semiconductors.
{\it Phys.\ Rev.\ Lett.\ \/}{\bf 68}, 2664-2667 (1992).
\item Ohno, H.  et al. (GaMn)As: A new diluted magnetic semiconductor based on 
GaAs. {\it Appl.\ Phys.\ Lett.\ \/} {\bf 69}, 363-365 (1996). 
\item Matsukura, F., Ohno, H., Shen, A. \& Sugawara, Y. Transport properties and origin of
ferromagnetism in (GaMn)As. {\it Phys.\ Rev.\ B \/}{\bf 57}, R2037-2040 (1998).
\item Reed, M.L.  et al. Room temperature ferromagnetic properties of (GaMn)N. 
{\it Appl.\ Phys.\ Lett.\ \/}{\bf 79}, 3473-3475 (2001).
\item Schlesinger, Z. et al. Unconventional charge gap formation in FeSi. 
{\it Phys.\ Rev.\ Lett.\ \/}{\bf 71}, 1748-1751 (1993).
\item van der Marel, D., Damascelli, A., Schulte, K.\ \& Menovsky, A.A.
Spin, charge, and bonding in transition metal mono-silicides. {\it Physica B\/} {\bf 244}, 138-147(1998).
\item Manyala, N. et al. Magnetoresistance from quantum interference
effects in ferromagnets. {\it Nature\/} {\bf 404}, 581-584 (2000).
\item Beille, J., Voiron, J.\ \& Roth, M. Long period helimagnetism in the
cubic-B20 Fe$_{1-x}$Co$_x$Si and Co$_x$Mn$_{1-x}$Si alloys. {\it Sol.\ St.\ Comm.\ \/}
{\bf 47}, 399-402 (1983). 
\item Wernick, J.H., Wertheim, G.K.\ \& Sherwood, R.C.
Magnetic behavior of the monosilicides of 3d-transition elements.
{\it Mat.\ Res.\ Bull.\ \/}{\bf 7}, 1431-1441 (1972). 
\item Aeppli, G.\ \& Fisk, Z. Kondo Insulators. {\it Comments Condens.\ 
Matter Phys.\ \/} {\bf 16}, 155-165 (1992). 
\item Jaccarino, V., Wertheim, G.K., Wernick, J.H., Walker, L.R. \& Arajs, S. Paramagnetic
excited states of FeSi. {\it Phys.\ Rev.\ \/} {\bf 160}, 476-482 (1967).
\item DiTusa, J.F., Friemelt, K., Bucher, E., Aeppli, G.\ \& Ramirez, A.P. Metal--insulator
transitions in Kondo insulator FeSi and classic semiconductors are similar. 
{\it Phys.\ Rev.\ Lett.\ \/} {\bf 78}, 2831-2834 (1997) and 
{\bf 78}, 4309(E) (1997). 
\item Chernikov, M.A.\ et al. Low temperature transport, optical, magnetic, and thermodynamic
properties of Fe$_{1-x}$Co$_x$Si. {\it Phys.\ Rev.\ B\/} {\bf 56}, 1366-1375 (1997). 
\item See e.g. Moriya, T. {\it Spin fluctuations in itinerant electron
magnetism.\/}, edited by Fulde, P.\ (Springer--Verlag, Berlin, 1985).
\item Mena, F.P.\ et al. Heavy carriers and non-Drude optical conductivity in MnSi.{\it Phys. Rev. B}
{\bf 67}, 241101 1-4 (2003).
\item Pfleiderer, C., Julian, S.R.\ \& Lonzarich, G.G. Non-Fermi-liquid nature of the
normal state of itinerant-electron ferromagnets. {\it Nature} {\bf 414}, 427-430 (2001).
\item See e.g., Campbell, I.A.\  \& Fert, A. {\it Ferromagnetic Materials.\/},
Edited by Wohlfarth, E.P.\ (North-Holland Pub.\ Co., Amsterdam, 1982), Vol.\ {\bf 3}.
\item Maranzana, F.E. Contribution to the theory of the anomalous Hall effect in
ferro- and antiferromagnetic materials. {\it Phys.\ Rev.\ \/}{\bf 160}, 421-429 (1967). 
\item Mattheiss, L.F.\ \& Hamann, D.R. Band Structure and Semiconducting Properties of FeSi. 
{\it Phys. Rev. B}, {\bf 47} 13114-13119 (1993).
\item Cattaneo, E. Hall--effect of Ce intermetallic compounds. {\it Z.\ Phys.\ B\/} {\bf 64}, 305-316 (1986). 
\item Coleman, P., Anderson, P.W. \& Ramakrishnan T.V. Theory of the anomalous Hall 
constant of mixed--valence systems. {\it Phys.\ Rev.\ Lett.\ \/}
{\bf 55}, 414-417 (1985).
\item Matl, P. et al. Hall effect of the colossal magnetoresistance manganite
La$_{1-x}$Ca$_x$MnO$_3$. {\it Phys.\ Rev.\ B \/} {\bf 57}, 10248-10251 (1998).
\item Otto M.J. et al. Half-metallic ferromagnets: II. Transport properties of NiMnSb
and related inter-metallic compounds. {\it J.\ Phys.\ Condens.\ Matter\/} {\bf 1}, 2351-2360 (1989).
\item Tomioka, Y. et al. Magnetic and electronic properties of a single crystal of ordered
double perovskite Sr$_2$FeMoO$_6$. {\it Phys.\ Rev.\ B \/} {\bf 61}, 422-427 (2000).
\item Jungwirth, T., Niu, Q. \& MacDonald, A.H. Anomalous Hall effect in ferromagnetic
semiconductors. {\it Phys.\ Rev.\ Lett.\ \/}
{\bf 88}, 207208-1 - 4 (2002).
\item Luttinger, J.M. Theory of the Hall effect in ferromagnetic substances.
{\it Phys.\ Rev.\ \/} {\bf 112}, 739-751 (1958).
\item Kadowaki, K.\ \& Woods, S.B. Universal relationship of the resistivity and specific--heat in 
heavy-Fermion compounds. {\it Sol.\ St.\ Comm.\ \/} {\bf 58}, 507-509 (1986).
\item Ye, J.\  et al. Berry phase theory of the anomalous Hall effect: application to colossal 
magnetoresistance manganites. {\it Phys.\ Rev.\ Lett.\ \/} {\bf 83}, 3737-3740 (1999).
\item Emel'yanenko, O.V., Lagunova, T.S., Nasledov, D.N.\ \& Talalakin, G.N. Formation
and properties of an impurity band in n-type GaAs. {\it Fiz.\ Tver.\ Tela \/} {\bf 7}, 
1315-1323 (1965).
\item Field, S.B \& Rosenbaum, T.F. Critical behavior of the Hall conductivity
at the Metal-Insulator Transition. {\it Phys.\ Rev.\ Lett.\ \/} {\bf 55}, 
522-524 (1985).
\item Chapman, P.W., Tufte, O.N., Zook, J.D.\ \& D. Long. Electrical properties of heavily doped
silicon. {\it J.\ Appl.\ Phys.\ \/} {\bf34}
3291-3294 (1963).
\item Leighton, C.,Terry, I. \& Becla, P. Metallic conductivity near the metal insulator transition
in Cd$_{1-x}$Mn$_x$Te. {\it Phys.\ Rev.\ B \/}
{\bf 58}, 9773-9782 (1998).
\item Carter, G.C.\ \& Pugh, E.M. Hall effect and transverse magnetoresistance in
some ferromagnetic iron--chromium alloys. {\it Phys.\ Rev.\ \/}{\bf 152}, 498-504 (1966).
\item Dorleijn, J.W. Electrical conduction in ferromagnetic metals. {\it Philips Res.\ 
Repts.\ \/} {\bf 31}, 287-410 (1976).
\item Lavine, J.M. Extraordinary Hall-effect measurements on Ni, some 
Ni alloys,and ferrites. {\it Phys.\ Rev.\ \/}{\bf 123}, 1273-1277 (1961).
\item McGuire, T.R., Gambino, R.J.\ \& Taylor, R.C. Hall effect in amorphous thin--film 
magnetic alloys. {\it J.\ Appl.\ Phys.\/}
{\bf 48}, 2965-2970 (1977).
\item Canedy, C.L., Gong, G.Q., Wang, J.Q.\ \& Xiao, G. Large magnetic hall effect in 
ferromagnetic Fe$_x$Pt$_{100-x}$ thin films. {\it J.\ Appl.\ Phys.\ \/} {\bf 79}, 6126-6128 (1996).
\item Lin, S.C.H. Hall effect in an amorphous ferromagnetic alloy. {\it J.\ Appl.\ Phys.\ \/}{\bf 40}, 
2175-2176 (1969). 
\item Bergmann, G.\ \& Marquardt, P. Resistivity of amorphous ferromagnetic Fe$_c$Au$_{1-c}$ alloys: Anisotropy
and field dependence. {\it Phys.\ Rev.\ B \/} {\bf 18}, 326-337 (1978). 
\item Pakhomov, A.B., Yan, X.\ \& Zhao, B. Observation of giant Hall effect in granular
magnetic films.  {\it J.\ Appl.\ Phys.\ \/} {\bf 79}, 
6140-6142 (1996). 
\item  Pakhomov, A.B., Yan, X.\ \& Zhao, B. Giant Hall effect in percolating ferromagnetic 
granular metal--insulator films. {\it Appl.\ Phys.\ Lett.\ \/} {\bf 67}, 
3497-3499 (1995). 
\item Samoilov, A.V, Beach, G., Fu., C.C., Yeh, N.-C. \& Vasquez R.P. Magnetic percolation 
and giant spontaneous Hall effect in La$_{1-x}$Ca$_x$CoO$_3$ ($0.2\le x \le 0.5$).
{\it Phys.\ Rev.\ B\/} {\bf 57}, R14032-14035 (1998). 
\item Katsufuji, T., Hwang, H.Y. \& Cheong, S-W. Anomalous Magnetotransport properties of
R$_2$Mo$_2$O$_7$ near the magnetic phase boundary. {\it Phys.\ Rev.\ Lett.\ \/}
{\bf 84}, 1998-2001 (2000).
\item Taguchi, Y. \& Tokura, Y. Magnetotransport phenomena in a metallic ferromagnet of the 
verge of a Mott transition: Sm$_2$Mo$_2$O$_7$. {\it Phys.\ Rev.\ B\/} {\bf 60}, 10280-10283 (1999).
\item Batlogg, B. et al. Charge dynamics in (LaSr)CuO$_4$ - from underdoping to overdoping.  
{\it J.\ Low Temp.\ Phys.\  \/} {\bf 95}, 23-31 (1994).
\item Bucher, B. et al. Charge dynamics of Ce--based compounds: Connection between the mixed
valent and Kondo--insulator states.  {\it Phys.\ Rev.\ B \/} {\bf 53}, R2948-2951 (1996).
\item Lapierre, F.\ et al. Hall effect in heavy Fermion systems. 
{\it J.\ Mag.\ Mag.\ Mat.\ \/} {\bf 63 \& 64}, 338-340 (1987).
\item Hadzic-Leroux, H.\ et al. Hall effect in heavy--Fermion systems: UPt$_3$, UAl$_2$, CeAl$_3$,
CeRu$_2$Si$_2$. {\it Europhys.\ Lett.\ \/} {\bf 1}, 579-584 (1986). 
\item Gao, Q.Z. et al. On the crystal field and Kondo effect in CeIn$_3$.
{\it J.\ Mag.\ Mag.\ Mat.\ \/} {\bf 47 \& 48}, 69-71 (1985). 
\item Djerbi, R.\  et al. Influence of Y and La alloying on the anomalous Hall-effect of 
Ce$_2$Ru$_2$Si$_2$. {\it J.\ Mag.\ Mag.\ Mat.\ \/} {\bf 76 \& 77}, 267-268(1988). 
\item Kasuya, T.\ et al.\ in {\it Valence Fluctuations in Solids.\/}, Falicov, L.M.,
Henke, W.\ \& Maple, M.B.\  (eds.) North-Holland Pub.\ Co.\  (Amsterdam), 281-284 (1981).
\item Penney, T.\ et al.\ Hall effect in the heavy Fermion systems CeCu$_6$ and UBe$_{13}$.
{\it J.\ Mag.\ Mag.\ Mat.\ \/} {\bf 54 - 57}, 370-372 (1986).

\end{enumerate}
\end{quote}

\clearpage

\noindent {\bf Fig. 1.} Phase diagram of Fe$_{1-x}$Mn$_x$Si and Fe$_{1-y}$Co$_y$Si.
(a)paramagnetic metallic (PMM), paramagnetic insulating (PMI) ($d\sigma / dT > 0$), 
and helimagnetic metallic (HMM)
phases at zero field are displayed. (b) Conductivity ($\sigma$)
at T = 2 K and H = 0 T vs.\ nominal Mn and Co concentration
($x$, $y$). (c) Lattice constant vs.\ $x$ and $y$
determined from powder X-ray diffraction measurements. 
(d) Carrier density as determined from Hall effect (for $H \geq 3$T) 
at 5.0 K vs.\ $x$ and $y$. The line 
is a single carrier per Mn or Co atom behavior.
(e) Hall mobility ($\mu_H = R_0 /\rho_{xx}$) as determined from the line
in (d) and zero field $\sigma$ measurements at 5.0 K 
vs.\ $x$, $y$. The blue bullet is 
$\mu_H$ for MnSi where the high field Hall effect gives an apparent 
$n = 1.7\times 10^{23}$ cm$^{-3}$ large enough to 
exceed the scale in frame $d$.

\noindent {\bf Fig. 2.} Comparison of Fe$_{0.9}$Co$_{0.1}$Si, Fe$_{0.1}$Mn$_{0.9}$Si, MnSi, and 
(Ga$_{1-z}$Mn$_z$)As ($z=0.053$) taken from Refs.\ 2 \& 3.
(a) Hall resistivity ($E_y / J_x$) of Fe$_{0.9}$Co$_{0.1}$Si at 5 K (blue circles), 15 K (black squares),
and 25 K (purple diamonds),  Fe$_{0.1}$Mn$_{0.9}$Si at 5 K (red triangles), MnSi at 5 K (green bullets),
and (GaMn)As at 2 K (orange line).
(b) Resistivity at zero field of Fe$_{0.9}$Co$_{0.1}$Si (blue circles),
Fe$_{0.1}$Mn$_{0.9}$Si (red triangles), MnSi (green bullets),
and (GaMn)As (orange line).
(c) Magnetization (symbols the same as in (a)).
(d) Magnetic susceptibility at 50 G (symbols the same as (b)).

\noindent {\bf Fig. 3.} Temperature and magnetic field dependence of the anomalous Hall effect.
(a) Low field Hall resistivity ($\rho_{xy}$) vs.\ field for several Fe$_{1-y}$Co$_y$Si and 
Fe$_{1-x}$Mn$_x$Si samples. 
$y=0.2$ sample is a single crystal. Symbols same as in (b). 
Inset: High field $\rho_{xy}$ of our $y=0.3$ sample.  
(b) Anomalous Hall constant ($R_S$) vs.\ temperature for several Fe$_{1-x}$Mn$_x$Si and
Fe$_{1-y}$Co$_y$Si samples as well as (Ga$_{1-z}$Mn$_z$)As ($z=0.053$) taken from Refs.\ 2 \& 3.
Open symbols indicate $T_c$ and
lines are guides to the eye.

\noindent {\bf Fig. 4.}  Hall effect of paramagnetic metals and insulators, and ferromagnetic metals at
1 kG and low temperature.
(a) Hall resistivity ($\rho_{xy}$)
at 1 kG vs.\ carrier concentration ($n$) at $\sim$5 K
except where noted below. Line is $\rho_{xy} = H/nec$, the Drude form. 
Data taken from the literature include GaAs$^{29}$, 
Ge:Sb$^{30}$, 
Si:P, Si:B$^{31}$, 
(InMn)As$^{1}$,
(GaMn)As$^{2}$,
Cd$_{0.92}$Mn$_{0.08}$Te$^{32}$,
dilute Fe Alloys$^{33}$, 
dilute Ni Alloys$^{34,35}$, 
CoMnSb, NiMnSb$^{23}$,
amorphous Co$_{0.72}$Gd$_{0.15}$Mo$_{0.11}$ films,
amorphous Co$_{0.70}$Gd$_{0.19}$Au$_{0.10}$ films (77 K)$^{36}$, 
Fe$_x$Pt$_{100-x}$ thin film$^{37}$, 
amorphous FeCP$^{38}$,
amorphous Fe$_{0.5}$Au$_{0.5}$$^{39}$, 
(NiFe)$_x$(SiO$_2$)$_{1-x}$ thin films$^{40}$, 
Ni$_x$(SiO$_2$) thin films$^{41}$,
La$_{0.7}$Ca$_{0.3}$CoO$_3$ (20 K)$^{42}$,
La$_{0.7}$Ca$_{0.3}$MnO$_3$ (290 to 100 K)$^{22}$,
Gd$_2$Mo$_2$O$_7$$^{43}$,
Sm$_2$Mo$_2$O$_7$$^{44}$, 
La$_{2-x}$Sr$_x$CuO$_4$ ($x=0.15,0.2$)$^{45}$,
Sr$_2$FeMoO$_6$$^{24}$,
CeBe$_{13}$$^{20}$,
UPt$_3$,
CePd$_3$ and Ce(Pd$_{1-x}$Ag$_x$)$_3$$^{20,46}$,
CeAl$_3$$^{47,48}$,
CeSn$_3$$^{20}$,
CeIn$_3$$^{49}$,
CeAl$_2$$^{20}$,
CeRu$_2$Si$_2$$^{50}$,
CeB$_6$$^{51}$,
CeCu$_6$$^{52}$,
and UAl$_2$$^{47,48}$.
Lines connect data at different $T$s for materials with strong $T$ dependencies.
(b) $\rho_{xy}$ at 1 kG vs.\ resistivity ($\rho_{xx}$)
at 5 K. 
(c) Anomalous Hall constant ($R_S$) vs.\ $\rho_{xx}$
at 5 K. Line is a $R_S \propto \rho_{xx}^2$
behavior. Symbols with arrows denote an upper limit.

\noindent {\bf Fig. 5.} Hall conductivity ($\sigma_{xy}$) of ferromagnetic metals and heavy 
Fermion materials.
Symbols and references are as in Fig.\ 4, ie.\ collected at 1 kG and 5 K except where noted
otherwise. Small dark--orange symbols are $\sigma_{xy}$ for Fe$_{1-y}$Co$_y$Si
for $5 < T < 75$ K and $0.05 < H < 5$ T with $y=0.01$ (bullets), $y=0.15$ (solid--triangles),
$y=0.2$ (+), and $y=0.3$ (solid--diamonds). Small blue asterisks are the (GaMn)As data 
for $5 < T < 120$ K from ref.\ {\it 2\/}. Red line is 
$\sigma_{xy} \propto M$ demonstrating the leading $M$ dependence of $\sigma_{xy}$
for itinerant magnets.  Purple line is a $1/ M^3$ dependence 
predicted from a simple phenomenology of heavy Fermion materials. A 
decreasing $\sigma_{xy}$ with $M$ is
most likely due to the stronger dependence of $\rho_{xx}$ 
than $M$ (for fixed $H$)
on the effective mass in these compounds. We note that for HFs
$\sigma_{xy} = \rho_{xy} / \rho_{xx}^2 \propto
\chi_p H / A^2 T^4$, where A is the coefficient of the Fermi liquid 
$T^2$ term in $\rho_{xx}$ and $\chi_p$
is the enhanced Pauli susceptibility. The Kadowaki-Woods relation sets 
$A \propto \gamma^2 \propto \chi_p^2$ 
where $\gamma$ 
is the coefficient of the linear $T$ dependent term in the electronic specific heat
and we have assumed a constant Wilson ratio ($\chi_p / \gamma$)$^{27}$.
Thus we expect $\sigma_{xy}\propto 
\chi_p / \chi_p^4\propto 1/M^3$ at low fields corresponding to the purple line.

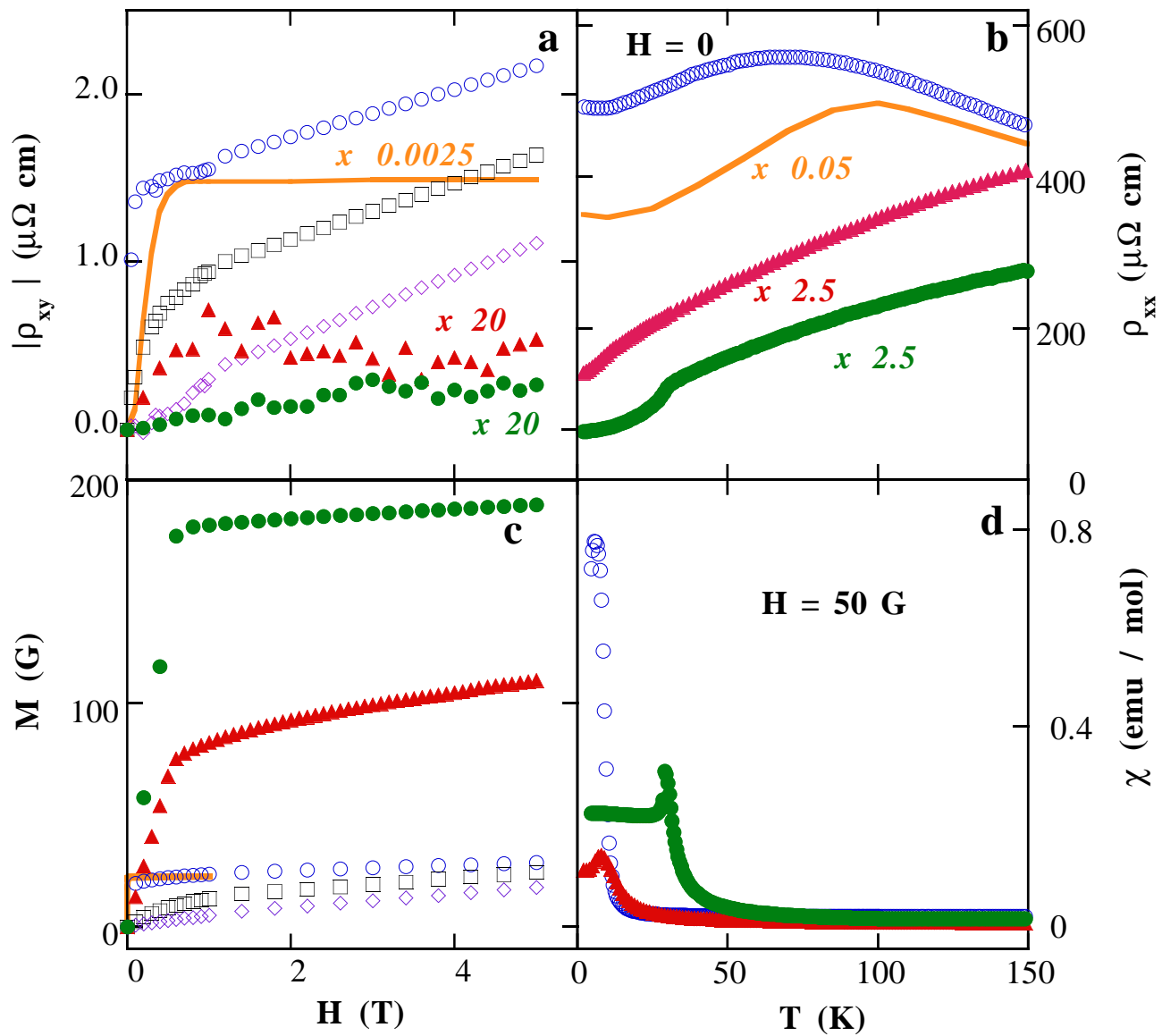
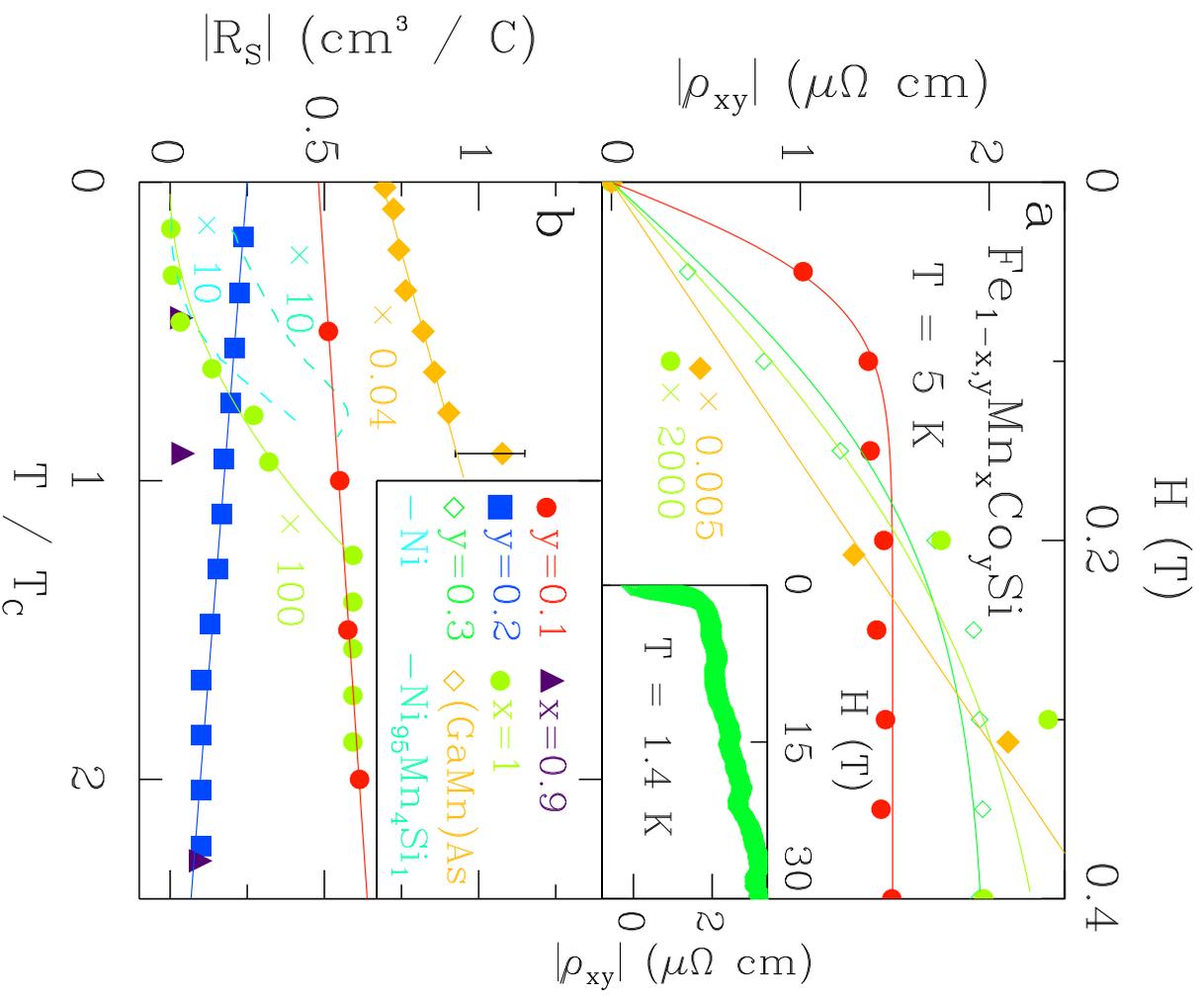
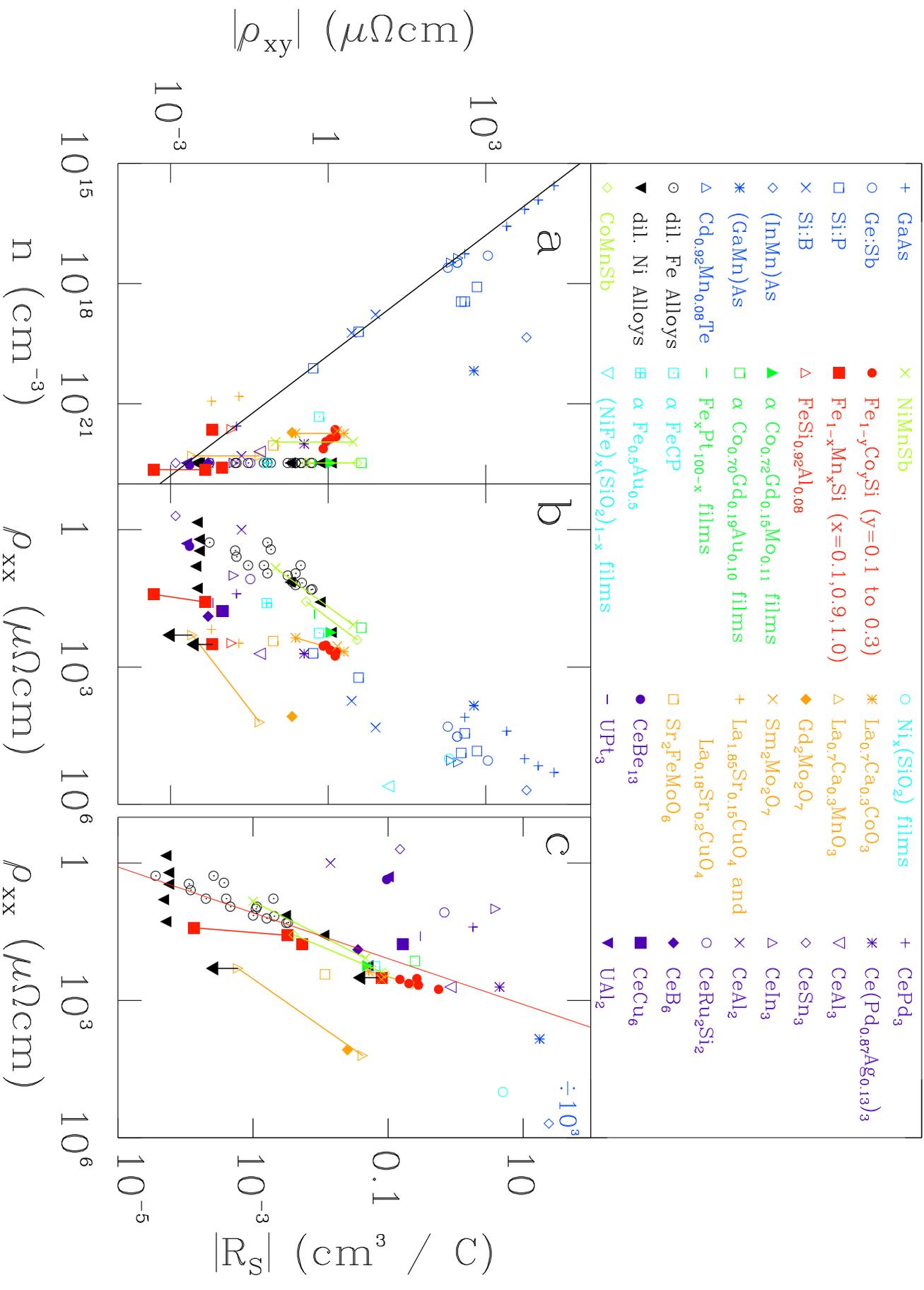
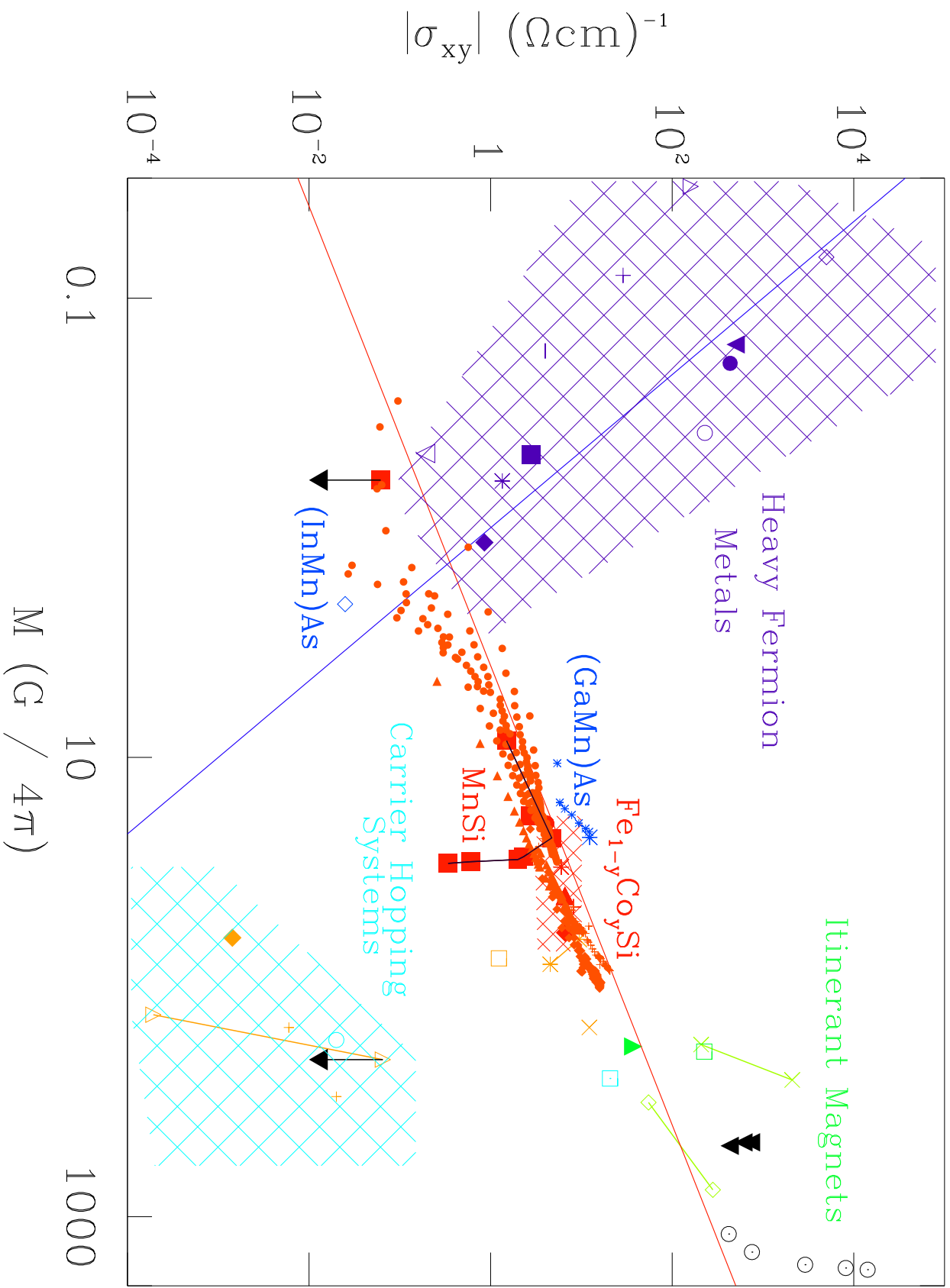
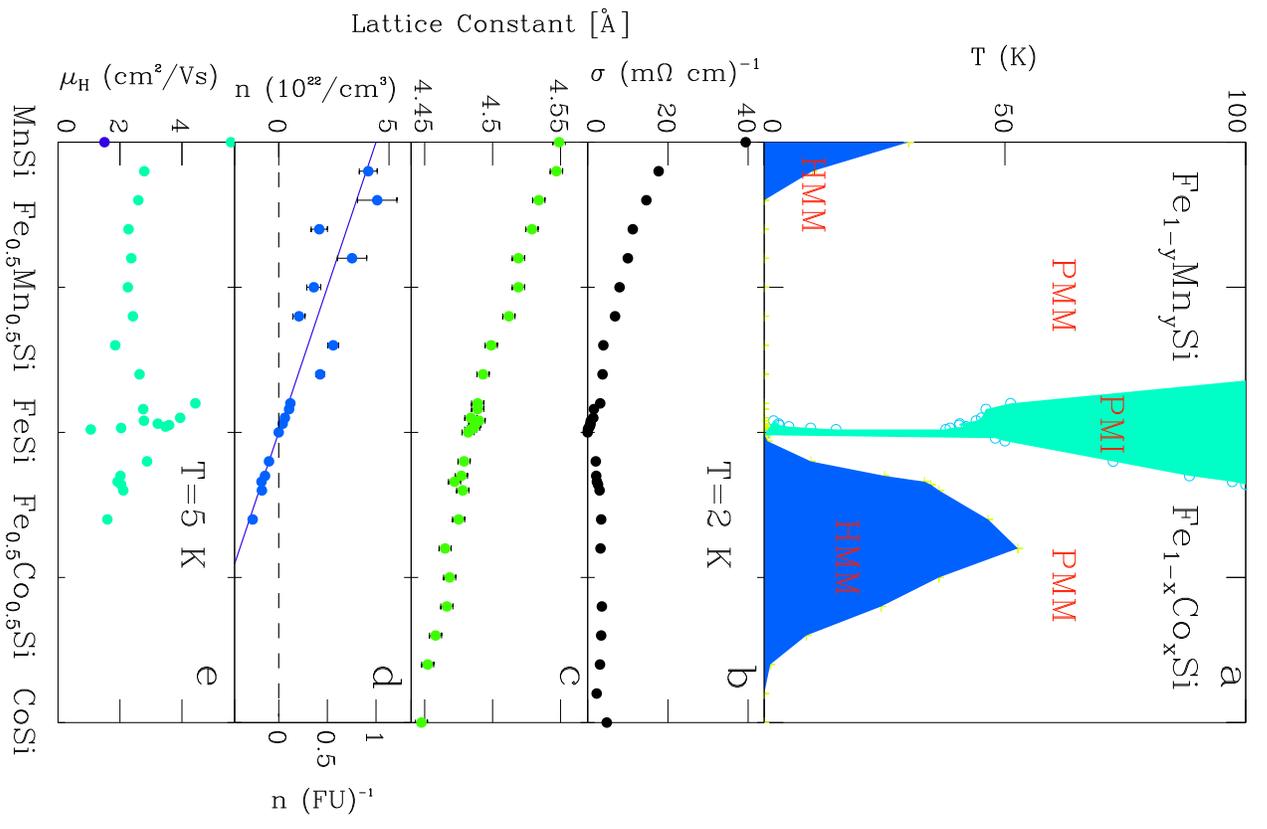
\end{document}